**Incommensurate magnetism in the coupled spin tetrahedra system $Cu_2Te_2O_5Cl_2$**

O. Zaharko*[1], H. M. Ronnow[1], A. Daoud-Aladine[1], S. Streule[1], F. Juranyi[1], J. Mesot[1]
H. Berger[2]
P. J. Brown[3]

[1]*Laboratory for Neutron Scattering, ETHZ & PSI, CH-5232 Villigen PSI, Switzerland*
[2]*Institute de Physique de la Matière Complexe, EPFL, CH-1015 Lausanne, Switzerland*
[3]*Institute Laue-Langevin, 156X, F-38042 Grenoble Cèdex, France*

PACS: 75.30-m, 61.12.Ld, 75.10.Jm

**Abstract**

Neutron scattering studies on powder and single crystals have provided new evidences for unconventional magnetism in $Cu_2Te_2O_5Cl_2$. The compound is built from tetrahedral clusters of S=1/2 $Cu^{2+}$ spins located on a tetragonal lattice. Magnetic ordering, emerging at $T_N$=18.2 K, leads to a very complex multi-domain, most likely degenerate, ground state, which is characterized by an incommensurate (ICM) wave vector **k**~[0.15, 0.42,1/2]. The $Cu^{2+}$ ions carry a magnetic moment of 0.67(1) $\mu_B$/ $Cu^{2+}$ at 1.5 K and form a four helices spin arrangement with two canted pairs within the tetrahedra. A domain redistribution is observed when a magnetic field is applied in the tetragonal plane ($H_c$≈0.5 T), but not for H||c up to 4 T. The excitation spectrum is characterized by two well-defined modes, one completely dispersionless at 6.0 meV, the other strongly dispersing to a gap of 2 meV.
The reason for such complex ground state and spin excitations may be geometrical frustration of the $Cu^{2+}$ spins within the tetrahedra, intra- and inter-tetrahedral couplings having similar strengths and strong Dzyaloshinski-Moriya anisotropy. Candidates for the dominant intra- and inter-tetrahedral interactions are proposed.

**Introduction**

Reduced dimensionality, geometrical frustration and low spin values often lead to quantum fluctuations resulting in interesting new ground states and spin dynamics [1]. The most famous examples are based on triangular units (triangular and kagomé lattices [2]) in two dimensions (2D) and tetrahedral clusters (FCC and pyrochlore lattices [3]) in 3D. The nature of the ground state in such systems is a subject of current strong interest, especially for the extreme quantum mechanical case S=1/2.
The copper tellurate $Cu_2Te_2O_5X_2$ (X= Cl, Br, space group P-4) [4] belong to a new family of such compounds. Their structure can be viewed as a stacking of layers of $Cu_4O_8Cl_4$ clusters along *c*. The four $Cu^{2+}$ ions within such a cluster, Cu1 (x, y, z), Cu2 (1-x, 1-y, z), Cu3 (y, 1-x, -z) and Cu4 (1-y, x, -z), form an irregular tetrahedron with two longer (Cu1-Cu2, Cu3-Cu4) and four shorter edges. The tetrahedra have a 2D square arrangement within the *ab*-layers.

The $Cu_2Te_2O_5Cl_2$ system attracted much attention as the first magnetic susceptibility data fitted well a model of isolated tetrahedra [4]. The observed broad maximum between 20 K - 30 K and a rapid drop at lower temperatures indicated a strength of the intra-tetrahedral coupling J~38.5 K. Raman spectroscopy, however, indicated a substantial inter-tetrahedral coupling along the *c* axis [5,6] and the data analysis have been performed [7] in terms of a dimerized model with the two pairs of $Cu^{2+}$ spins: Cu1-Cu2 and Cu3-Cu4. Further magnetic susceptibility and specific heat measurements [5] indicated an onset of antiferromagnetic (AF) order at $T_N$=18.2 K, thus confirming the importance of inter-tetrahedral couplings. Recent neutron diffraction studies [8] revealed the details of the magnetic order: it is incommensurate and very complex.

In spite of considerable progress in experimental studies, the relevant intra- and inter-tetrahedral magnetic interactions in $Cu_2Te_2O_5Cl_2$ remain a puzzle due to the rather complex 3D exchange topology (Fig. 1). The intra-tetrahedral spin interactions are mediated via the superexchange paths Cu-O-Cu and can be described by $J_1$ and $J_2$ exchange constants. It was suggested [9] according to the Goodenough rules, that the $J_2$ interaction should be weakly antiferromagnetic if not ferromagnetic, while $J_1$ is antiferromagnetic and rather weak [9]. From the band structure calculations [10] it is expected that within the *ab*-layers the $J_a$ and $J_d$ inter-tetrahedral couplings are substantial (see Fig. 1) and mediated by the halogen *p* orbitals. Between the adjacent layers the tetrahedra interact through the super-superexchange paths Cu-O…O-Cu and the corresponding inter-tetrahedral couplings are as well important.

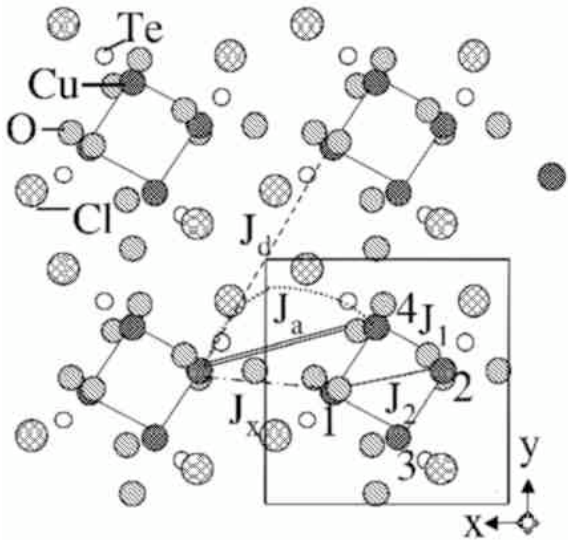

Fig. 1. The xy-projection of the $Cu_2Te_2O_5Cl_2$ crystal structure (z=-1/2 ÷1/2) with the $J_1$, $J_2$ and possible inter-tetrahedral exchange paths [10]. The double-line segment is normal to the incommensurate component of the wave vector **k'** [8].

The nature of the magnetic ordering transition is also still unclear. A strong spin-lattice coupling near $T_N$ has been suggested from magnetic susceptibility and thermal conductivity studies [11]. However, the infrared spectroscopy study [12], as well as high-resolution neutron diffraction study [8] gave no evidence of any lattice distortion.

We present neutron diffraction and inelastic scattering results, which supply new information on the magnetic ground state, spin-orbit coupling and spin dynamics. We hope this will provide a better starting point for theoretical modeling [13-18].

**Experimental**

High-purity powder and single crystals of $Cu_2Te_2O_5Cl_2$ were prepared by the halogen vapor transport technique, using $TeCl_4$ and $Cl_2$ as transport agents. Neutron powder diffraction (NPD) patterns were collected in the temperature range 1.5 K - 30 K on the DMC instrument, with a neutron wavelength of λ=4.2 Å and on the high-resolution HRPT instrument (λ=1.889 Å) at SINQ, Villigen, Switzerland.

The neutron single crystal diffraction (NSCD) experiments on two crystals of dimensions 2.5 x 3 x15 mm$^3$ and 2 x 3.5 x 6 mm$^3$ were carried out using the diffractometers TriCS at SINQ (λ=1.18 Å) and D15 (λ=1.17 Å) at the high-flux ILL reactor, France. The NSCD with applied magnetic field were performed on a larger 1 cm$^3$ crystal on TriCS (λ=1.18 Å and 2.4 Å) for three field directions H||$a$, $b$ and $c$.

The inelastic neutron scattering experiment was carried out on 15 g powder on neutron time-of-flight spectrometer FOCUS (λ=2.8 Å and 4 Å) at SINQ and on a large 1.5 cm$^3$ crystal on the IN8 thermal neutron three-axis spectrometer at ILL. IN8 was configured with doubly focusing monochromator and analyzer with a 5 cm graphite filter in $k_f$ and fixed final neutron energy of 14.7 meV.

**Results**

**Diffraction**

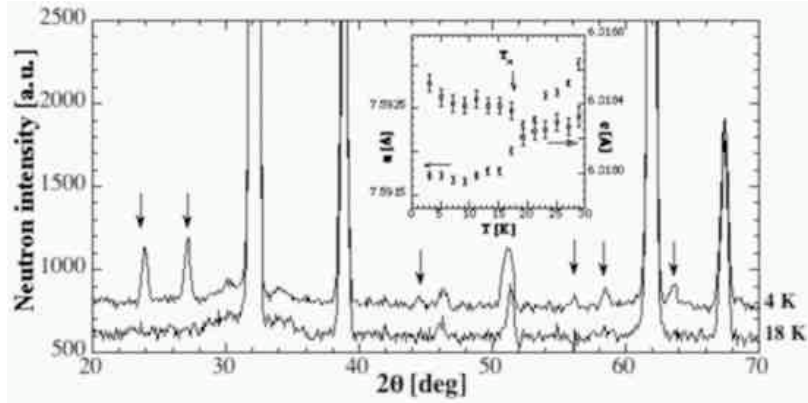

Fig. 2. The 18 K and 4 K DMC NPD patterns of $Cu_2Te_2O_5Cl_2$ (λ=4.2 Å), arrows point to magnetic reflections. Inset: temperature evolution of the lattice constants from HRPT NPD data (λ=1.889 Å).

Below $T_N$=18 K tiny magnetic peaks appeared (Fig. 2) in DMC neutron diffraction patterns of $Cu_2Te_2O_5Cl_2$, as already reported in [8]. Moreover, the onset of magnetic order at $T_N$ can be followed from the structural peaks in the HRPT data. The lattice constant $a(b)$ significantly decreases with temperature above $T_N$ and changes very little below $T_N$ (Fig. 2 inset). This implies that the spin-lattice coupling is substantial, but no changes of the crystal structure could be determined from the neutron patterns.

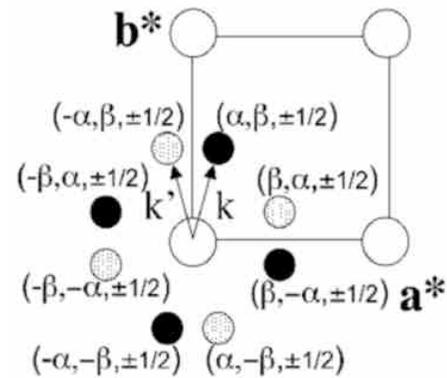

To determine correctly the magnetic ground state it is very important to elucidate the magnetic symmetry. The [001] projection of neutron diffraction pattern of a typical $Cu_2Te_2O_5Cl_2$ single crystal is presented in Fig. 3.

Fig. 3. The [001] projection of the reciprocal space of a $Cu_2Te_2O_5Cl_2$ single crystal. The black circles correspond to the **k** magnetic reflections and the dotted circles – to the **k'** set; α=0.15, β=0.422.

Up to 16 magnetic satellites of the (0,0,0) reflection have been observed. The reflection set denoted by black circles arises from four arms of the star of the incommensurate (ICM) wave vector **k**~[0.150, 0.422,1/2]. The reflections denoted by dotted circles correspond to the star of another wave vector **k'**~[-0.150,0.422,1/2]. The **k** and **k'** vectors are not related by the symmetry elements of the group P-4, and could belong to growth crystallographic twins. Since for several studied crystals the **k'** reflections are absent, we conclude that the **k** and **k'** sets are independent. Interestingly, the intensity ratio between the two first magnetic reflections for the **k** and **k'** sets is different, implying that the magnetic structures associated with these two stars are different.

We further tried to clarify if the magnetic structure is single-k or multi-k. In the case of a single-k magnetic structure the **k**(**k'**) set contains contributions of four configuration domains. The configuration domains must all have the same structure but possibly different populations. Each of them could have two 180 deg domains and two chiral domains, which cannot be distinguished from nonpolarized neutron diffraction experiment. In the case of multi-k structure, the four arms of the star build one magnetic structure.

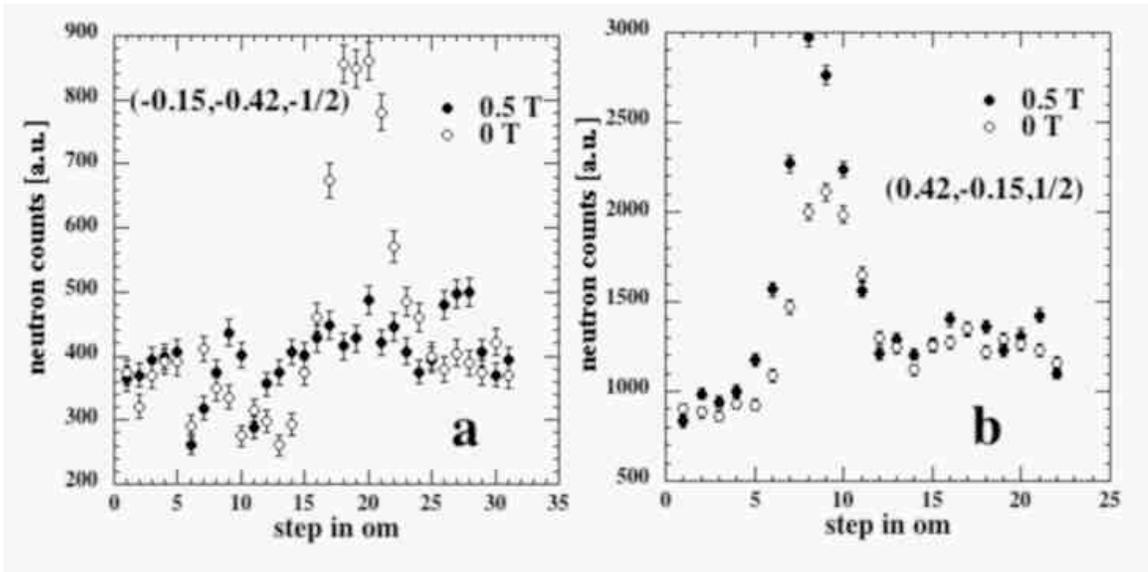

Fig. 4. The TriCS ω-scans of the (-0.15,-0.42,-1/2) (a) and (0.42,-0.15,1/2) (b) magnetic reflections (λ=2.4 Å) at H=0 T and H=0.5 T along *a*.

It is possible to distinguish the single-k or multi-k cases by studying the variation of magnetic intensities as a function of an applied magnetic field H. We performed such study for fields along *a*, *b* and *c* crystal directions. For H along *c* the intensities of the magnetic reflections *hkl* with *l*>0 increase and for *l*<0 they decrease, but all reflections persist up to 4 T. This implies a change of the spin arrangement with respect to the zero-field magnetic structure, but without domain redistribution and/or meta-magnetic transition. For H||*a* a transition occurs at $H_c$~0.5 T: intensities of the (±α,±β,±l/2) reflections vanish, while the magnetic reflections of the type (±β,±α,±l/2) change their intensities smoothly (Fig. 4).

Flipping the field to –*a* results in the same behaviour. Switching off the field restores partly the vanished peaks. A similar picture is observed for H||*b*, but now the (±β,±α,±l/2) reflections vanish at ~0.5 T. Our results imply that 0.5 T applied in the tetragonal plane is

enough to depopulate the domains with the propagation vector nearly normal (90±10 deg) to the field direction. This supports the idea that the magnetic structure is a single-k and not a multi-k structure.

The model for the **k'** magnetic structure has been recently elaborated in [8]. The only symmetry constraint is imposed by the commensurate component of the wave vector. It implies that the *ab*-layers of tetrahedra alternating along *c* carry oppositely oriented spins. The magnetic moments of the four $Cu^{2+}$ ions in the unit cell can be independent and a generalized helix characterizing spin arrangement of each of them in the crystal is expressed as

$\mathbf{S_j} = \mathbf{A} \cos(\mathbf{k} \cdot \mathbf{r_j} + \psi) + \mathbf{B} \sin(\mathbf{k} \cdot \mathbf{r_j} + \psi)$.

The spin components are modulated by the wave vector **k**, $\mathbf{r_j}$ is the radius vector to the origin of the *j*-th unit cell. **A** and **B** are orthogonal vectors, which define the magnitude and direction of the axes of the helix, whilst $\psi$ defines its phase. For the most general case 27 independent parameters should be considered. Since the available number of experimental observations does not allow to refine with confidence all of them, we imposed physically sound constraints: identical moment value for all four independent $Cu^{2+}$ ions and a circular envelope of the helices. This lowered the number of independent parameters to 12.

Very good fits were obtained for a model (Fig. 5) in which the 4 $Cu^{2+}$ moments in each tetrahedron form two canted pairs: Cu1-Cu2 and Cu3-Cu4. The two spins of the pair share a common (**A**, **B**) plane, but the associated helices have different phases $\psi$. The difference between the phases defines the canting angle between spins of the pair $\alpha$. The canting angle for the first pair is $\alpha_{12}=38(6)$ deg and for the second pair $\alpha_{34}=111(14)$ deg. The amplitude of the magnetic moment carried by each $Cu^{2+}$ ion is 0.67(1) $\mu_B$ at 1.5 K. It is interesting that the vector sum of the spins of one pair is the same for all tetrahedra in the crystal ($m_{12}=1.27(6)$ $\mu_B$, $m_{34}=0.76(14)$ $\mu_B$), whilst the local magnetic moment of the tetrahedra is not zero and changes from one unit cell to another.

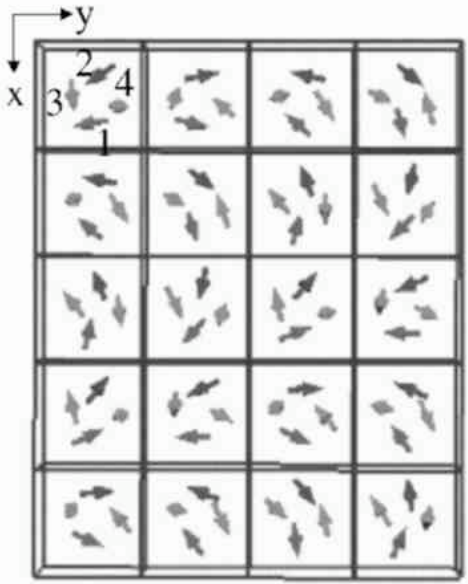

Fig. 5. The xy-projection of the $Cu_2Te_2O_5Cl_2$ magnetic structure with the spin tetrahedra at the z=0 layer.

For an isotropic exchange the spin state of the tetrahedra would be zero, so our model might suggest that the two intra-tetrahedral couplings $J_1$ and $J_2$ are different and that the $J_2$ interaction is stronger.

Such a particular magnetic ground state might be a consequence of one dominant interaction or result from contributions of several inter-tetrahedral couplings of similar strengths. We tried to attribute the observed reciprocal wave vector **k'**=(-0.15,0.42,1/2) to some specific exchange inter-tetrahedral path in the structure and found a simple correlation not to **k'**, but to the (-0.15,-0.58,0) vector. As this vector is related by a lattice translation to the (-0.15,0.42,0), the two vectors mean a different choice of the unit cell of the same magnetic structure. The ICM component is orthogonal to a specific set of planes containing the $Cu^{2+}$ ions.

One of these planes, presented by a double-line segment in Fig. 1, passes through the Cu2-Cu4 ions of the adjacent tetrahedra. This corresponds to the $J_a$ coupling, which according to Ref. [10] is substantial and is mediated by the halogen orbitals. Based on these considerations we suggest that $J_a$ is the dominant inter-tetrahedral coupling.

**Inelastic scattering**

Several inelastic neutron scattering studies have been performed to investigate the excitation spectrum of the $Cu_2Te_2O_5Cl_2$ system. Spectra from powder revealed in the ordered phase below $T_N$ a spherically averaged density of states extending to a maximum at 6 meV, above which no significant scattering was detected up to 15 meV (fig. 6). Raising the temperature above $T_N$, spectral weight shifted downwards to a broad quasi-elastic peak.

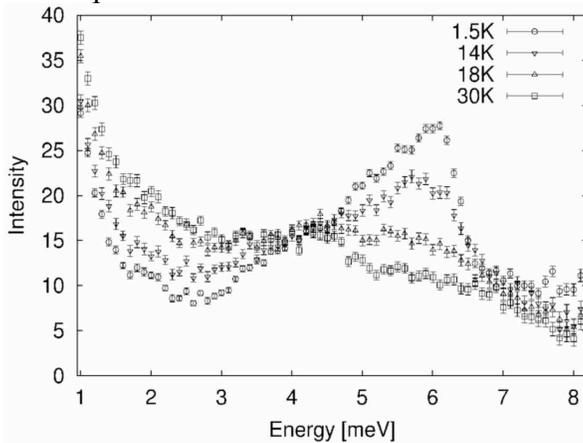

Fig. 6. Inelastic powder neutron scattering of $Cu_2Te_2O_5Cl_2$ integrated between 0.8 Å$^{-1}$ and 3.3 Å$^{-1}$ for four temperatures between 2 K and 30 K. With incident energy $E_i$=10.4 meV inelastic focusing was adjusted to provide optimal resolution at 5 meV of 0.45 meV (full-width-half-maximum). For single crystal neutron spectroscopy a single crystal was aligned with (3,1,0) and (0,0,1) in the horizontal scattering plane, such that (0.42,0.15,1/2) and the equivalent magnetic Bragg peaks were accessible given the wide (~ 5 deg) vertical resolution. Scans performed along the Q=(h,h/3,3/2), (0.45,0.15,l) and (0,0,l) directions revealed two well defined excitation modes. One mode is completely dispersionless at 6.0 meV and shows no variation in intensity as a function of Q. The other mode displays strong dispersion along both accessible directions from a maximum energy close to the flat mode down to a minimum energy gap of 2.1 meV at the same positions in Q as the ICM magnetic Bragg peaks.

While more detailed modeling of the excitation spectrum is under way, several important conclusions can be read directly off the figure. If the system would be a collection of very weakly coupled tetramers, one would expect a series of essentially dispersionless modes. The strong dispersion observed in our measurements imply strong inter-tetrahedral coupling both within the *ab*-plane and along the *c*-axis. Secondly, a classical ordered magnetic structure with continuous symmetry of the order parameter should have gapless spin waves emerging from the magnetic Bragg peaks. The rather large energy gap must involve strong anisotropy terms in the Hamiltonian, whose origin remains to be determined.

**Summary**

The presented results of neutron diffraction and inelastic neutron scattering evidence new details of the magnetic ordering in the $Cu_2Te_2O_5Cl_2$ compound. The idea of a single-k magnetic structure is strongly supported by the observed magnetic domains redistribution in an applied magnetic field. The presence of two different **k** and **k'** structures suggests that a number of ground states with equal or close energies might exist. The discovered relation between the incommensurate component of the wave vector and the inter-tetrahedral coupling $J_a$ invites for a theoretical revision of the $Cu_2Te_2O_5X_2$ system. The peculiar features of the spin excitation spectrum deserve further study.

**Acknowledgments**

The work was performed at SINQ, Paul Scherrer Institute, Villigen, Switzerland and at ILL reactor, Grenoble, France. We thank Prof. F. Mila, Dr. M. Prester and Prof. A. Furrer for fruitful discussions and Swiss NCCR research pool MANEP of the Swiss NSF for financial support.

**Literature**

1. *Quantum magnetism*, *Lect. Notes Phys.*, **645**, *Springer-Verlag, Berlin, Heidelberg (2004)*.
2. F. Mila, *Eur. J. Phys.* **21**, *499 (2000)*.
3. A. P. Ramirez, *Geometrical frustration, in Handbook of Magnetic Materials*, **13**, *Elsevier Science (2001)*.
4. M. Johnsson, K. W. Törnroos, F. Mila, P. Millet, *Chem. Mater.* **12**, *2853 (2000)*.
5. P. Lemmens, K.-Y. Choi, E. E. Kaul, C. Geibel, K. Becker, W. Brenig, R. Valenti, C. Gross, M. Johnsson, P. Millet, F. Mila, *Phys. Rev. Lett.* **87**, *227201 (2001)*.
6. C. Gross, P. Lemmens, M. Vojta, R. Valenti, K.-Y. Choi, H. Kageyama, Z. Hiroi, N. M. Mushnikov, T. Goto, M. Johnsson, P. Millet, *Phys. Rev. B* **67**, *174405 (2003)*.
7. J. Jensen, P. Lemmens, C. Gross, *Europhys. Lett.* **64**, *689(2003)*.
8. O. Zaharko, A. Daoud-Aladine, S. Streule, J. Mesot, P.-J. Brown, H. Berger, *Phys. Rev. Lett.* **93**, *217206 (2004)*.
9. M. H. Whangbo, H. J. Koo, D. Dai, *J. Solid State Chem.* **176**, *417 (2003)*.
10. R. Valenti, T. Saha-Dasgupta, C. Gros, H. Rosner, *Phys. Rev. B* **67**, *245110 (2003)*.
11. M. Prester, A. Smontara, I. Zivkovi'c, A. Bilusi'c, D. Drobac, H. Berger, F. Bussy, *Phys. Rev. B* **69**, *180401R (2004)*.
12. A. Perucchi, L. Deiorgi, H. Berger, P. Millet, *Eur. Phys. J. B* **38**, *65(2004)*.
13. W. Brenig, K. W. Becker, *Phys. Rev. B* **64**, *214413 (2001)*.
14. W. Brenig, *Phys. Rev. B* **67**, *64402 (2003)*.
16. K. Totsuka, H. - J. Mikeska, *Phys. Rev. B* **66**, *54435 (2002)*.
17. V. N. Kotov, M. E. Zhitomirsky, M. Elhajal, F. Mila, *Phys. Rev. B* **70**, *214401 (2004)*.
18. V. N. Kotov, M. E. Zhitomirsky, M. Elhajal, F. Mila, *J. Phys.: Condens. Matter* **16**, *S905 (2004)*.